\documentclass[aps,prl,floatfix,twocolumn,superscriptaddress,preprintnumbers,nofootinbib]{revtex4-1}

\usepackage{textcomp} 
\usepackage{slashed}
\usepackage{epsfig,latexsym,cancel,amssymb,amsmath,verbatim,mathrsfs}
\usepackage{color}
\usepackage{graphicx}
\usepackage{enumitem}
\usepackage[colorlinks,citecolor=blue]{hyperref}

\newcommand{\beq}{\begin{equation}}
\newcommand{\eeq}{\end{equation}}
\newcommand{\bea}{\begin{eqnarray}}
\newcommand{\eea}{\end{eqnarray}}
\newcommand{\met}{\not{\!\!{\rm E}}_{T}}
\newcommand{\nn}{\nonumber}
\newcommand{\tttt}{t\bar{t}t\bar{t}}
\newcommand{\tth}{t\bar{t}H}

\allowdisplaybreaks[1]

\begin{document}

\title{Probing Higgs Width and Top Quark Yukawa Coupling from $t\bar{t}H$ and $t\bar{t}t\bar{t}$ Productions}

\author{Qing-Hong Cao}
\email{qinghongcao@pku.edu.cn}
\affiliation{Department of Physics and State Key Laboratory of Nuclear Physics and Technology, Peking University, Beijing 100871, China}
\affiliation{Collaborative Innovation Center of Quantum Matter, Beijing 100871, China}
\affiliation{Center for High Energy Physics, Peking University, Beijing 100871, China}

\author{Shao-Long Chen}
\email{chensl@mail.ccnu.edu.cn}
\affiliation{Key Laboratory of Quark and Lepton Physics (MoE) and Institute of Particle Physics, Central China Normal University, Wuhan 430079, China}
\affiliation{Center for High Energy Physics, Peking University, Beijing 100871, China}

\author{Yandong Liu}
\email{ydliu@pku.edu.cn}
\affiliation{Department of Physics and State Key Laboratory of Nuclear Physics and Technology, Peking University, Beijing 100871, China}

\begin{abstract}
We demonstrate that four top-quark production is a powerful tool to constrain the top Yukawa coupling. The constraint is robust in the sense that it does not rely on Higgs boson decay. Taking into account the projection of the $t\bar{t}H$ production by the ATLAS collaboration, we obtain a bound on Higgs boson width, $\Gamma_H\leq 3.1~\Gamma_H^{\rm SM}$, at the 14 TeV LHC with an integrated luminosity of $300~{\rm fb}^{-1}$. Increasing  the luminosity to $500~{\rm fb}^{-1}$ yields $\Gamma_H\leq 2.1~\Gamma_H^{\rm SM}$. 

\end{abstract}

\maketitle

Four years after the Higgs boson discovery we still know little about Higgs boson width ($\Gamma_H$) and its couplings to fermions in the Standard Model (SM). For its smallness the Higgs boson width cannot be measured directly from the line-shape of Higgs boson resonance. One way to determine $\Gamma_H$ is through the $gg\to H\to ZZ$ channel by comparing the production rate in the vicinity of Higgs resonance with the rate away from the resonance~\cite{Caola:2013yja}. So far only an upper bounds are obtained; for example, the current bounds on $\Gamma_H$ at 95\% confidence level are $\Gamma_H \leq (4.5\sim 7.5)\times \Gamma^{\rm SM}_H$ by the ATLAS collaboration~\cite{Aad:2015xua} and $\Gamma_H \leq 5.4~\Gamma^{\rm SM}_H$ by the CMS collaboration~\cite{Khachatryan:2014iha}. Similarly, the top Yukawa coupling ($y_{Ht\bar{t}}$) is not directly measured yet, although the Higgs boson discovery indicates the Higgs boson must interact with top quarks to generate Higgs-gluon-gluon effective coupling. The top Yukawa coupling can be measured in the rare $\tth$ production on condition that the Higgs boson decays exactly as in the SM. Precise information of Higgs boson width and top Yukawa coupling will help us to decipher Higgs boson property and also shed light on new physics beyond the SM. 
In this work we discuss the measurement of $\Gamma_H$ and $y_{Ht\bar{t}}$ in the four top quark ($\tttt$) production and the $\tth$ production at the Large Hadron Collider (LHC). We demonstrate that the combination of the two production channels imposes stringent bounds on $\Gamma_H$ and $y_{Ht\bar{t}}$. 

As reported by the ATLAS collaboration~\cite{ATL-PHYS-PUB-2014-016}, the top Yukawa coupling could be measured in the $\tth$ production with 
an ultimate precision of about 20\% at the 14~TeV LHC with an integrated luminosity ($\mathcal{L}$) of $300~{\rm fb}^{-1}$. 
Under the narrow width approximation the production cross section of $pp\to \tth \to t\bar{t}xx$ is 
\bea
&&\sigma(pp\to \tth \to t\bar{t}xx) \nn\\
&=& \sigma^{\rm SM}(pp\to \tth\to t\bar{t} xx)\times \kappa_t^2\kappa_x^2\frac{\Gamma_{H}^{\rm SM}}{\Gamma_{H}}\nn\\
&\equiv & \sigma^{\rm SM}(pp\to \tth\to t\bar{t} xx)\times\mu^{xx}_{\tth},
\eea
where $\kappa_t\equiv y_{Htt}/y_{Htt}^{\rm SM}$ and $\kappa_x\equiv y_{Hxx}/y_{Hxx}^{\rm SM}$ are the scaling factors of Higgs couplings. The signal strength $\mu^{xx}_{\tth}$, defined as 
\beq
\mu^{xx}_{\tth}\equiv \frac{\sigma}{\sigma^{\rm SM}}= \frac{\kappa_t^2 \kappa_x^2 }{R_\Gamma}
\qquad {\rm with}\qquad R_\Gamma\equiv \frac{\Gamma_H}{\Gamma_H^{\rm SM}},
\label{eq:tth}
\eeq
is expected to be measured with uncertainties~\cite{ATL-PHYS-PUB-2014-016}
\begin{align}
&\overline{\mu}_{\tth}^{\gamma\gamma} = 1.00\pm 0.38~,\quad 
&&\overline{\mu}_{\tth}^{ZZ} = 1.00\pm 0.49~, \nn\\
&\overline{\mu}_{\tth}^{\mu\mu} = 1.00\pm 0.74~,\quad 
&&\overline{\mu}_{\tth}^{\rm ~comb} = 1.00\pm 0.30~,
\label{eq:atlas}
\end{align}
at the 14~TeV LHC with $\mathcal{L}=300~{\rm fb}^{-1}$. 
Here $\overline{\mu}_{\tth}^{\rm ~comb}$ refers to the result of combining multiple Higgs decay modes. The $\kappa_t $, $\kappa_x$ and $\Gamma_H$ parameters in $\mu_{\tth}$ are independent, therefore, one cannot determine them from the $\tth$ production alone. Bounds on the $\kappa_t$, $\kappa_x$ and $R_\Gamma$ could be derived from a global analysis of various Higgs boson productions and decays~\cite{ATL-PHYS-PUB-2014-016}. Nevertheless it is still valuable to consider one specific channel to directly bound on the three parameters. Luckily, there is a large hierarchy among branching ratios of Higgs decay modes. That ensures us to consider two special cases:
\begin{itemize}[leftmargin=1.5em]
\item[i)] $\Gamma_H\simeq \Gamma_H^{\rm SM}$: it is a good approximation for the $H\to \mu^+\mu^-$ and $H\to\gamma\gamma$ modes because modifications on those rare decays would not affect the total width dramatically. One thus can determine the bound on the product of $\kappa_t$ and $\kappa_x$ as
\beq
\kappa_t^2\kappa_x^2 =\overline{\mu}_{\tth},
\label{eq:r1}
\eeq
assuming other couplings of the Higgs boson are the same as the SM predictions.
\item[ii)] $\kappa_x \simeq 1$: Higgs boson might decay into a pair of invisible particles and modify the total width. A bound on $\kappa_t$ and $R_\Gamma$ is 
\beq
\frac{\kappa_t^2}{R_\Gamma} = \overline{\mu}_{\tth}.
\label{eq:r2}
\eeq 
\end{itemize}
If the top-quark Yukawa coupling could be directly measured or constrained in one particular Higgs production channel, then one can impose bounds on $\kappa_x$ and $R_\Gamma$ based on Eqs.~\ref{eq:r1} and~\ref{eq:r2}, respectively. Below we show that the $\tttt$ production is a powerful tool to constrain the top Yukawa coupling. 

\begin{figure}[t]
\includegraphics[scale=0.33]{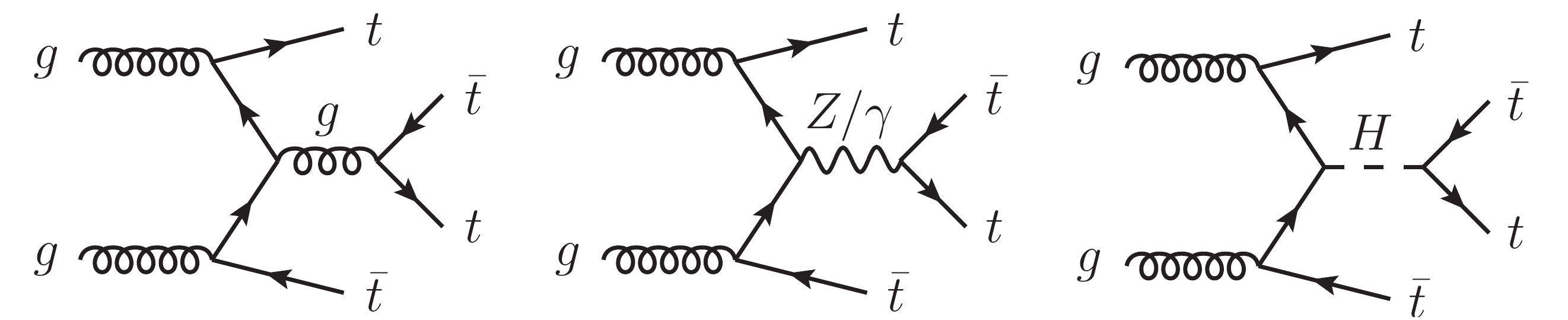}
\caption{Illustrative Feynman diagrams of $\tttt$ productions.}
\label{fig:feyn}
\end{figure}

Figure~\ref{fig:feyn} displays the representative Feynman diagrams of the $\tttt$ production, which occurs either through the gluon mediation, the electroweak gauge-boson mediation, or the Higgs boson mediation in the SM. We name the corresponding matrix elements as $\mathcal{M}_{g}$, $\mathcal{M}_{Z/\gamma}$, and $\mathcal{M}_{H}$.
There are two advantages of the Higgs-induced $\tttt$ production: i) no dependence on the Higgs boson width; ii) the cross section proportional to the top quark Yukawa coupling to the fourth power, i.e.  
\beq
\sigma(\tttt)_H \propto  \kappa_t^4 \sigma^{\rm SM}(\tttt)_H,
\eeq
where $\sigma^{\rm SM}(\tttt)_H$ denotes the SM production cross section. The not-so-small interferences among the three kinds of Feynman diagrams are also accounted. 
Since the QCD and electroweak gauge interactions of top quarks have been well established, we consider only the top Yukawa coupling might differ from the SM value throughout this work. As a result, the cross section of $\tttt$ production is 
\beq
\sigma(\tttt) = \sigma^{\rm SM}(\tttt)_{g+Z/\gamma} + \kappa_t^2 \sigma^{\rm SM}_{\rm int} + \kappa_t^4 \sigma^{\rm SM}(\tttt)_H,
\eeq
where
\bea
\sigma^{\rm SM}(\tttt)_{g+Z/\gamma}&~\propto~& \left|\mathcal{M}_{g} + \mathcal{M}_{Z/\gamma}\right|^2, \nn\\
\sigma^{\rm SM}(\tttt)_{H}&\propto& \left|\mathcal{M}_{H}\right|^2, \nn\\
\sigma^{\rm SM}(\tttt)_{\rm int} &\propto& \mathcal{M}_{g+Z/\gamma}\mathcal{M}^\dagger_{H}+\mathcal{M}^\dagger_{g+Z/\gamma}\mathcal{M}_{H}.
\eea
We use MadEvent~\cite{Alwall:2007st} to calculate the leading order cross section of  $\tttt$ production in the SM. The numerical results are summarized as follows:
\begin{align}
&                &\text{8~TeV}~~& & \text{14~TeV}~~& \nn\\
&\sigma^{\rm SM}(\tttt)_{g+Z/\gamma}: &1.193~{\rm fb}, & & 12.390~{\rm fb}, &\nn\\
&\sigma^{\rm SM}(\tttt)_H:  &0.166~{\rm fb},& & 1.477~{\rm fb},&\nn\\
&\sigma^{\rm SM}(\tttt)_{\rm int}: & -0.229~{\rm fb}, && -2.060~{\rm fb}.&  
\end{align}
The numerical results shown above are checked with CalcHEP~\cite{Belyaev:2012qa}. A high integrated luminosity is needed to reach a $5\sigma$ discovery of the rare $\tttt$ production. However, null searching results in the low luminosity operation of the LHC are also useful because they can be used to constrain the top Yukawa coupling. For example, a 95\% CL bound, $\sigma(\tttt)\leq 23~{\rm fb}$, is reported recently by the ATLAS~\cite{Aad:2015kqa} and the CMS collaborations~\cite{Khachatryan:2014sca} at the 8~TeV LHC. That yields a bound of $\kappa_t\leq 3.49$. The $\kappa_t$ bound, though loose, is robust in the sense that it does not depend on how the Higgs boson decays. 

Next we examine how well the top-quark Yukawa coupling could be measured in the $\tttt$ production at the future LHC. A special signature of the $\tttt$ events is the same-sign charged leptons (SSL) from the two same-sign top quarks.  The ATLAS and CMS collaborations have extensively studied the same sign lepton pair signal at the LHC~\cite{ATLAS:2013tma,Chatrchyan:2013fea}. The other two top quarks are demanded to decay hadronically in order to maximize the production rate. Therefore, the topology of the signal event consists of two same-sign charged leptons, four $b$-quarks, four light-flavor quarks, and two invisible neutrinos. In practice it is challenging to identify four $b$-jets. Instead, we demand at least 5 jets are tagged and three of them are identified as $b$-jets. The two invisible neutrinos appear as a missing transverse momentum ($\met$) in the detector. Thus,  the collider signature of interests to us is two same-sign leptons, at least five jets and three of them tagged as $b$-jets, and a large $\met$. 

The SM backgrounds for same-sign leptons can be divided into three categories: i) prompt same-sign lepton pair from SM rare process, including 
di-boson and $W^\pm W^\pm jj$; ii) fake lepton, which comes from heavy quark jet, namely $b$-decays, and the dominant one is the $t\bar{t}+X$ events~\cite{ATLAS:2012sna}; iii) charge misidentification. As pointed out by the CMS collaboration~\cite{Chatrchyan:2013fea}, the background from charge mis-identification is generally much smaller and stays below the few-percent level. We thus ignore this type of backgrounds in our simulation and focus on those non-prompt backgrounds $t\bar{t}+X$ and rare SM processes contributions. For four top quark production process another feature worthy being specified is that multiple $b$-jets decay from top quark appear in the final state. Same-sign lepton plus multiple $b$-jets has a significant discrimination with the backgrounds.
Another SM process can contribute the same-sign lepton are the di-boson production, however, it can be highly suppressed by the request of tagging multiple jets in the final state. Therefore, the major backgrounds are from the $t\bar{t}+X$ and $W^\pm W^\pm jj$ channels. 

Both the signal and background events are generated at the parton level using MadEvent~\cite{Alwall:2007st} at the 14 TeV LHC. The higher order QCD corrections are taken in accounts by multiplying the leading order cross sections with a next-to-leading-order $K$-factor, e.g., $K_{F}=1.27$ for the $\tttt$ production~\cite{Bevilacqua:2012em}, $K_{F}=1.4$ for the $\bar{t}t$ production~\cite{Ahrens:2011px, Czakon:2013goa}, $K_{F}=1.22$ for the $\bar{t}tW^+$ channel and $K_{F}=1.27$ for the $\bar{t}tW^-$ channel~\cite{Campbell:2012dh},  $K_{F}=1.49$ for the $\bar{t}tZ$ production~\cite{Lazopoulos:2008de,Garzelli:2011is,Kardos:2011na,Garzelli:2012bn,Maltoni:2015ena,Frixione:2015zaa}, and $K_{F}=0.9$ for the $W^\pm W^\pm jj$ channel~\cite{Jager:2009xx,Melia:2010bm}.  We use Pythia~\cite{Sjostrand:2014zea} to generate parton showering and hadronization effects. The Delphes package~\cite{deFavereau:2013fsa} is used to simulate detector smearing effects in accord to a fairly standard Gaussian-type detector resolution given by $\delta E/E= \mathcal{A}/\sqrt{E/{\rm GeV}}\oplus \mathcal{B}$,
where $\mathcal{A}$ is a sampling term and $\mathcal{B}$ is a constant term.  For leptons we take $\mathcal{A}=5\%$ and $\mathcal{B}=0.55\%$, and for jets we take $\mathcal{A}=100\%$ 
and $\mathcal{B}=5\%$.
We require the charged lepton has a transverse momentum $p^{\ell}_T$ greater than $20$ GeV, rapidity $\left|\eta_{\ell}\right|\leq 2.5~$ and its overlap with jets $\Delta R_{j\ell} = \sqrt{(\Delta \eta)^2 + (\Delta \phi)^2} \geq 0.4$. The $\met$ is then defined to balance the total transverse momentum of visible objects.

\begin{figure}
\centering
\includegraphics[scale=0.32]{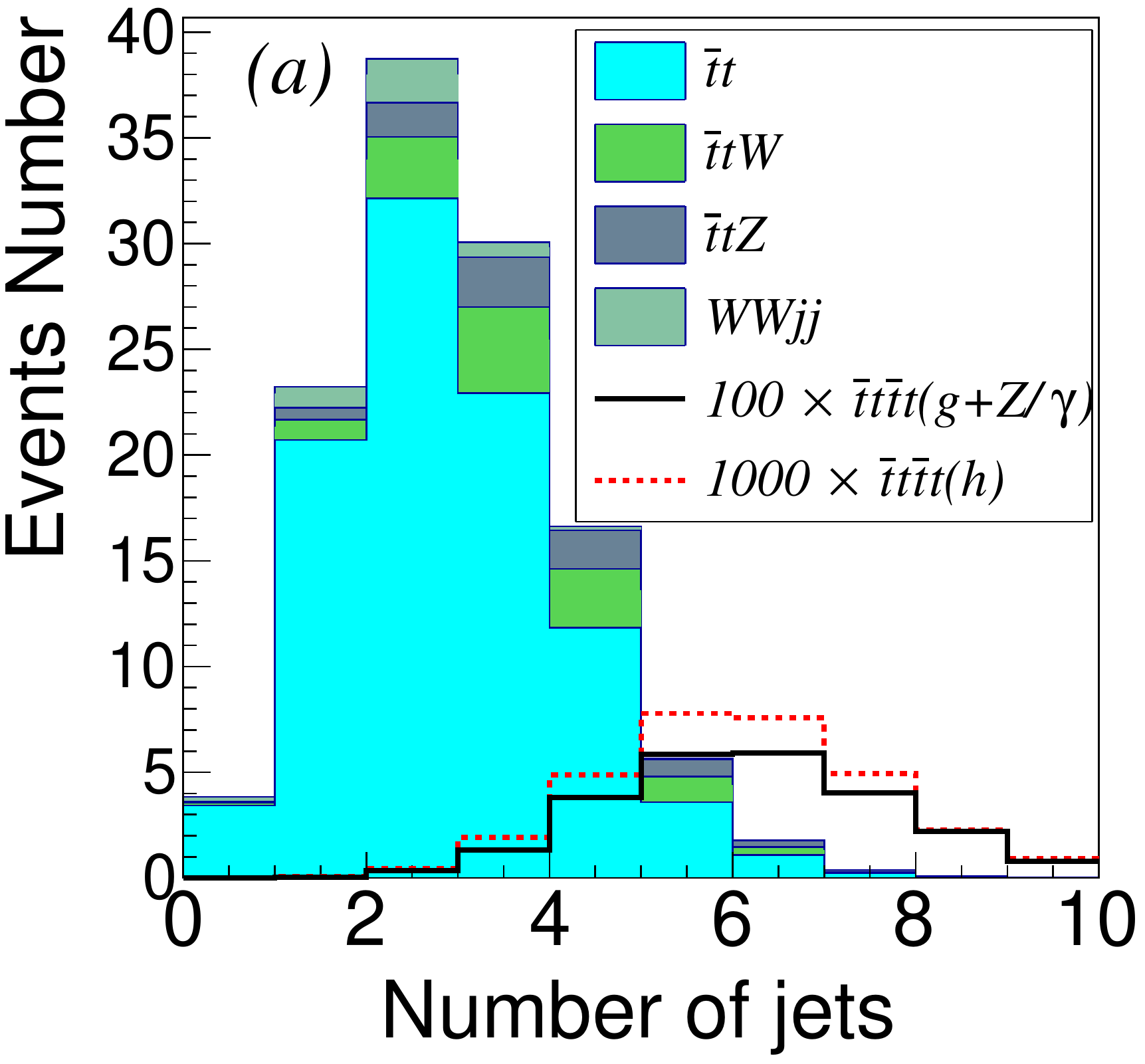}
\includegraphics[scale=0.32]{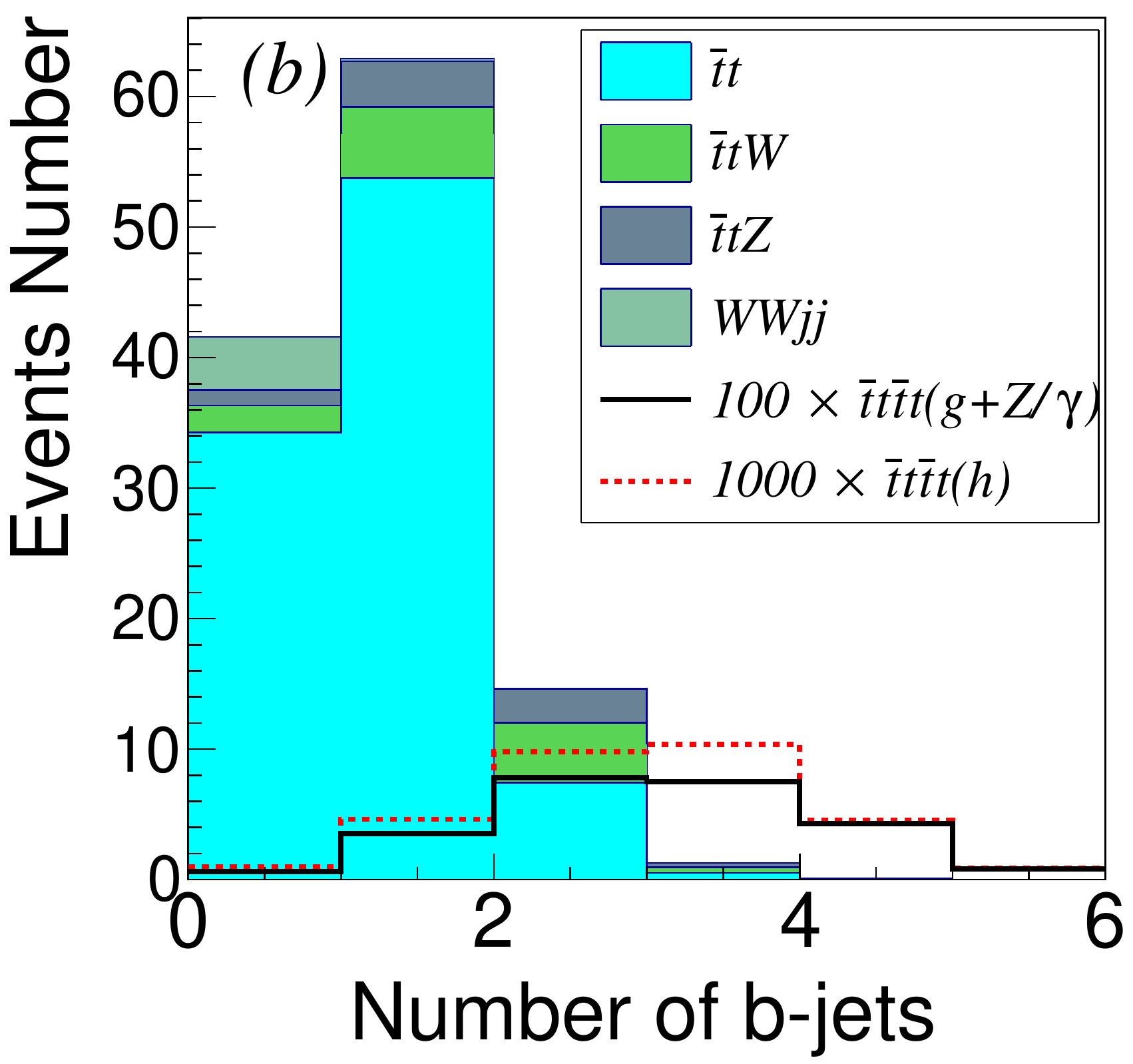}
\caption{The numbers of the reconstructed jets (a) and $b$-tagged jets (b) in the signal and background events at the 14~TeV LHC with an integrated luminosity of $1~{\rm fb}^{-1}$. To better character the signal distribution the cross section has been rescaled to 1000 times. No cuts except for same-sign lepton pair have been applied.}
\label{nbj}
\end{figure}

Figure~\ref{nbj} displays the numbers of reconstructed jets (a) and $b$-tagged jets (b) in the signal and background processes. It is clear that the signal event exhibits often five or more jets. Demanding at least three identified $b$-jets would efficiently reject those SM backgrounds. In the simulation we impose kinematics cuts listed as follows:
\begin{align}
& {\rm Basic:} && p_T^{j,\ell}\geq 20~{\rm GeV}, \quad |\eta^{j,\ell}|<2.5, &&\nn\\
& {\rm SSL:} && N_{\ell^\pm}=2, \nn\\
&{\rm Jets:}  && N_{\rm jets}\geq 5, \quad N_{b-{\rm jets}} \geq 3, \nn\\
& \met: &&\met\ge 100~{\rm GeV},\nn\\
& m_T: && m_T \geq 100~{\rm GeV}, \nn\\
& H_T: &&H_T \geq 700~{\rm GeV}. \label{eq:cuts}
\end{align}
Here $m_T$ denotes the transverse mass of the leading charged lepton ($\ell_1$) and the $\met$, defined as
\beq
m_T = \sqrt{2p_T^{\ell_1}\met (1 - \cos \Delta \phi)}, 
\eeq
where $\Delta\phi$ is the azimuthal angle between the $\ell_1$ lepton and the $\met$. The $m_T$ cut is to remove those backgrounds involving  leptonically decayed $W$ bosons. The $H_T$ is the  the scalar sum of the transverse momenta of all the visible particles and the missing energy $\met$.

\begin{table}
\caption{The numbers of the signal and background events at the 14 TeV LHC with an integrated luminosity of $300~{\rm fb}^{-1}$. The kinematics cuts listed in each row are applied sequentially.}
\label{tbl:14tev}
\begin{tabular}{c|c|c|c|c|c|c} \hline
&Basic &SSL&Jets&$\met$ &$m_T$&$H_T$ \\ \hline \hline
 $\bar{t}t\bar{t}t_H$ &562.74&9.37& 4.18&2.07&1.13& 0.94 \\ \hline
 $\bar{t}t\bar{t}t_{g+Z/\gamma}$&4720.59  &74.23&33.85&17.02 &10.12&8.78 \\ \hline
 $\bar{t}t\bar{t}t_{\rm int}$&-781.81&-12.84&-5.75&-3.11&-1.80&-1.75 \\ \hline
\hline
 $\bar{t}t$&$2.5\times 10^8$ & 28802.4 & 44.1 & 18.9 &0 &0 \\ \hline
 $\bar{t}tW^+$& 32670& 2359.5&36.9& 17.7& 12.3& 8.7 \\ \hline
 $\bar{t}tW^-$& 16758 & 1397.1&49.5 & 9.9& 4.5 &  4.5 \\ \hline
 $\bar{t}tZ$ &24516& 2309.4& 20.1 & 10.8&10.8& 9.3 \\ \hline
 $W^\pm W^\pm j j$ &4187.7& 1147.5& 0.11 &0&0&0 \\ \hline
\end{tabular}
\end{table}

Table~\ref{tbl:14tev} shows the numbers of the signal and the background events after a series of kinematics cuts at the 14~TeV LHC with an integrated luminosity of $300~{\rm fb}^{-1}$. The $\tttt$ production channels through the gluon, the electroweak gauge-boson and the Higgs boson mediation share similar kinematics, therefore, all the $\tttt$ production channels exhibit similar efficiencies for each cut shown in Table~\ref{tbl:14tev}. The major backgrounds in the SM are from the $t\bar{t}W^\pm$ and $t\bar{t}Z$ productions. About 22.5 background events remain after all the cuts. 

Next we discuss how well the top Yukawa coupling can be probed in the $\tttt$ production at the future LHC. As there are few events of both the signal and the backgrounds after the kinematics cuts, we obtain a $2\sigma$ exclusion limit on the $\tttt$ production rate using~\cite{Cowan:2010js}
\beq
\sqrt{-2\left[n_b \ln\left(\frac{n_s+n_b}{n_b}\right)-n_s\right]} =2
\label{eq:excl}
\eeq
where $n_s$ and $n_b$ are the numbers of signal and background events, respectively. If a null result is observed on top of the 22.5 background events, then the number of signal events cannot exceed 10.9, from which we obtain $\kappa_t\leq 1.6$ with $\mathcal{L}=300~{\rm fb}^{-1}$. Bounds for other integrated luminosities can be derived similarly, yielding $\kappa_t\leq 2.2$ for $\mathcal{L}=100~{\rm fb}^{-1}$ and $\kappa_t\leq 1.2$ for $\mathcal{L}=500~{\rm fb}^{-1}$.

\begin{figure}
\includegraphics[scale=0.4]{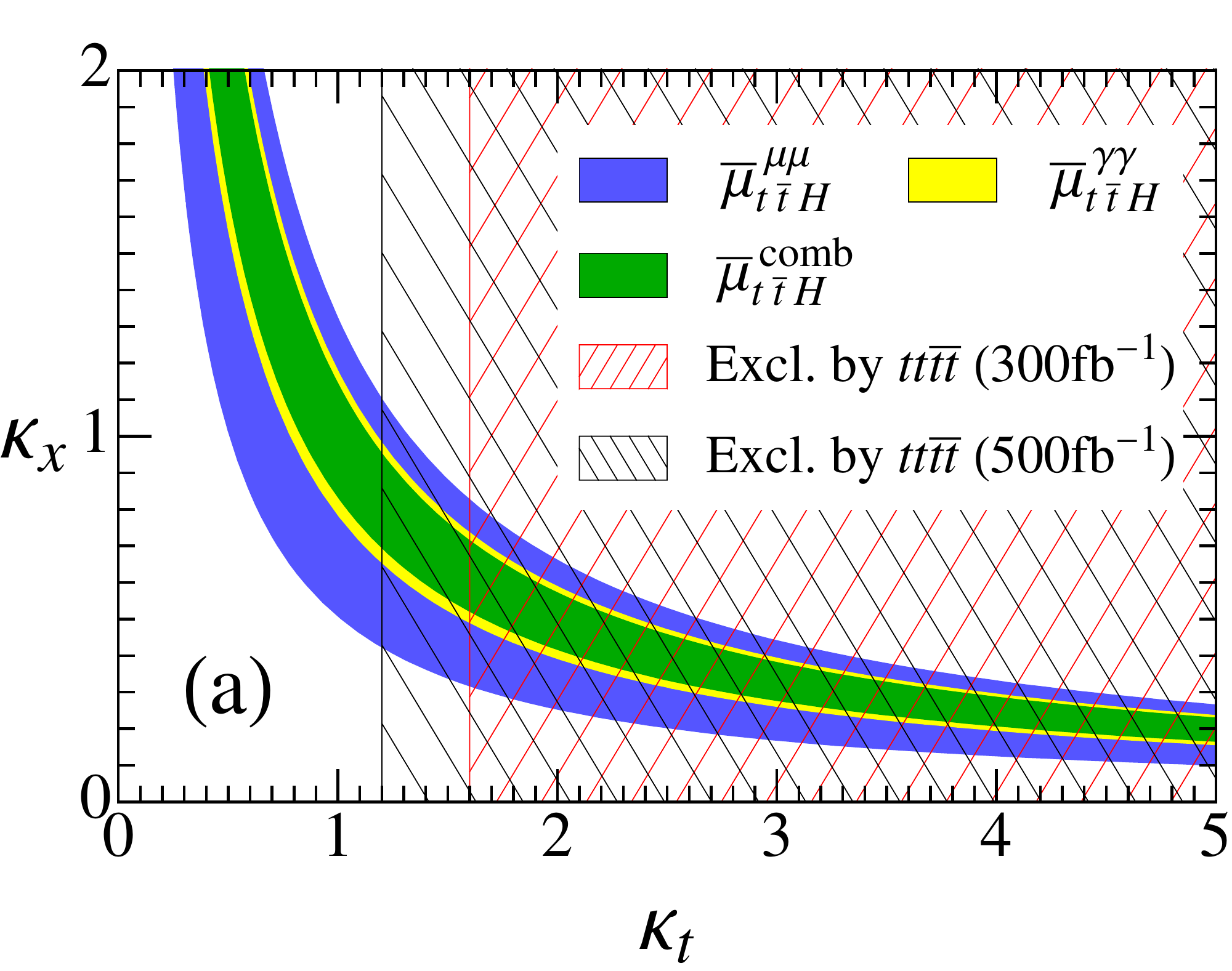}\quad
\includegraphics[scale=0.4]{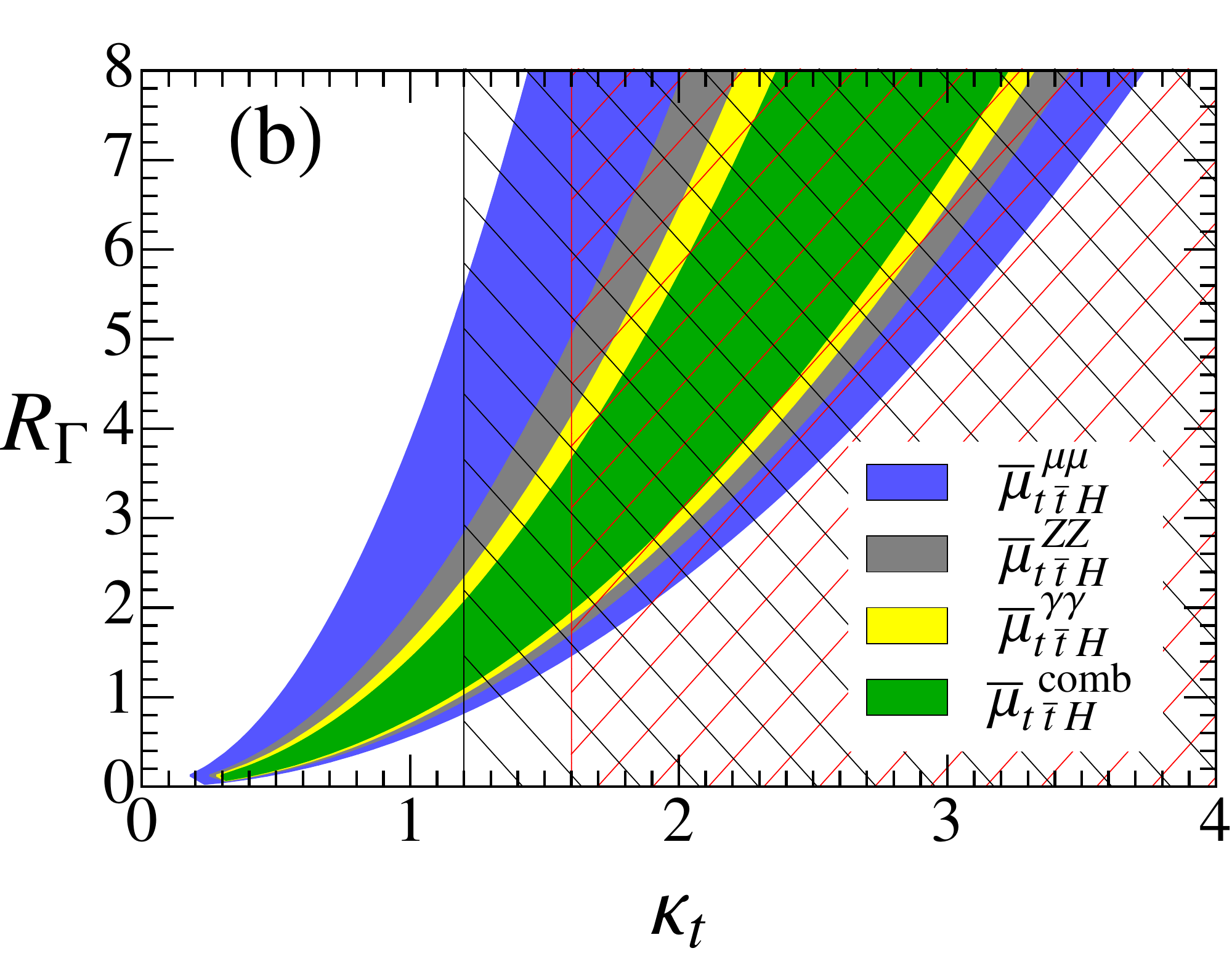}
\caption{The relative uncertainty on the signal strength $\mu_{\tth}$ projected in the plane of $\kappa_t$ and $\kappa_x$ (a) and in the plane of $\kappa_t$ and $R_\Gamma$ (b) at the 14 TeV with $\mathcal{L}=300~{\rm fb}^{-1}$ for $H\to \gamma\gamma$ (yellow), $H\to \mu^+\mu^-$ (blue), $H\to ZZ$ (gray), and also the combination (green). The red (black) meshed region is excluded by the $\tttt$ production  with $\mathcal{L}=300~(500)~{\rm fb}^{-1}$, respectively, if null signal events were observed.}  
\label{fig:limits}
\end{figure}

Taking into account the $\tth$ measurement projection at the 14 TeV LHC with $\mathcal{L}=300~{\rm fb}^{-1}$, one can derive a lower bound on $\kappa_x$ and an upper bound on $R_{\Gamma}$. Figure~\ref{fig:limits} shows the relative uncertainty on the signal strength $\mu_{\tth}$ projected in the plane of of $\kappa_t$ and $\kappa_x$ (a) and in the plane of $\kappa_t$ and $R_\Gamma$ (b). The blue band represents the $\tth$ measurement in the $H\to \mu^+\mu^-$ mode, the yellow band denotes the $H\to \gamma\gamma$ mode, and the gray band labels the $H\to ZZ$ mode. The green band is the result of combining different channels of Higgs production and decay. See Eq.~\ref{eq:atlas} for details. The red and black meshed regions are excluded by the $\tttt$ production with $\mathcal{L}=300~{\rm fb}^{-1}$ and $500~{\rm fb}^{-1}$, respectively, if null results were reported on top of the SM background.

First, we consider the correlation between $\kappa_t$ and $\kappa_x$ in the case of $\Gamma_H\simeq \Gamma_H^{\rm SM}$. In Fig.~\ref{fig:limits}(a) we plot constraints on rare Higgs-decay modes, $H\to \gamma\gamma$ (yellow) and $H\to \mu\mu$ (blue), assuming all Higgs couplings except the top Yukawa coupling the same as in the SM. The $\kappa_t$ exclusion limit derived from the $\tttt$ production requires that $\kappa_\mu\geq 0.32$ and $\kappa_{\gamma}\geq 0.49$ with $\mathcal{L}=300~{\rm fb}^{-1}$. Accumulating more luminosities improves the $\kappa_t$ bound and mildly tightens  the $\kappa_x$ bound, e.g., $\kappa_\mu\geq 0.43$ and $\kappa_{\gamma}\geq 0.66$ with $\mathcal{L}=500~{\rm fb}^{-1}$. The combination of multiple Higgs production channels yields a slightly tighter constraint. 

Secondly, consider all the Higgs couplings as in the SM, i.e. $\kappa_x=1$. We obtain the correlation between $\kappa_t$ and $R_\Gamma$ shown in Fig.~\ref{fig:limits}(b). The $\gamma\gamma$ ($ZZ$, $\mu^+\mu^-$) mode demands $R_\Gamma\leq 4.1$ (5.0, 9.8), respectively, at the 14 TeV LHC with $\mathcal{L}=300~{\rm fb}^{-1}$.  The combination analysis demands $\Gamma_H\leq 3.7~\Gamma_H^{\rm SM}$. Increasing the integrated luminosity to $500~{\rm fb}^{-1}$ leads to much tighter constraints on $R_\Gamma$ as follows: the $\gamma\gamma$ ($ZZ$, $\mu^+\mu^-$) mode demands $R_\Gamma\leq 2.3$ (2.8, 5.5), respectively. The combined analysis requires $\Gamma_H\leq 2.1~\Gamma_H^{\rm SM}$.

\medskip
The work is supported in part by the National Science Foundation of China under Grand No. 11175069, No. 11275009 and No. 11422545.

\bibliographystyle{apsrev}
\bibliography{reference}

\end{document}